\documentclass{article}
\usepackage[utf8]{inputenc}
\pdfoutput=1

\usepackage{hyperref}
\hypersetup{hidelinks}
\usepackage[square,numbers]{natbib}  %see https://www.overleaf.com/learn/latex/Bibliography_management_with_natbib
\usepackage{graphicx}
\usepackage{xspace}
\usepackage{xcolor}
\usepackage{subcaption}
\usepackage{amssymb,amsmath,amsthm}
\usepackage{algorithm,algorithmicx}
\usepackage[noend]{algpseudocode}
\usepackage{todonotes}
\usepackage{tikz,paralist,ifthen}  % For optproblem.

\graphicspath{{fig/}}

\newcommand{\eg}{\textit{e.g.}\xspace}
\newcommand{\ie}{\textit{i.e.}\xspace}

\newcommand{\maxneg}{\ensuremath{\mathsf{M}^-}\xspace}
\newcommand{\maxpos}{\ensuremath{\mathsf{M}^+}\xspace}
\newcommand{\minneg}{\ensuremath{\mathsf{m}^-}\xspace}
\newcommand{\minpos}{\ensuremath{\mathsf{m}^+}\xspace}
\newcommand{\inQmax}{\ensuremath{inQ_{\mathsf{M}}}\xspace}
\newcommand{\inQmin}{\ensuremath{inQ_{\mathsf{m}}}\xspace}
\newcommand{\Qmax}{\ensuremath{Q_{\mathsf{M}}}\xspace}
\newcommand{\Qmin}{\ensuremath{Q_{\mathsf{m}}}\xspace}

\newcommand{\adj}[2]{\mbox{($#1$~$#2$)}\xspace}

\newcommand{\rev}{\ensuremath{rev}\xspace}
\newcommand{\T}{\ensuremath{\mathcal{T}}\xspace}
\newcommand{\Tlt}[1]{\ensuremath{T_{<#1}}\xspace}

\newcommand{\Tgt}[1]{\ensuremath{T_{>#1}}\xspace}
\newcommand{\Tge}[1]{\ensuremath{T_{\geq#1}}\xspace}

\newcommand{\Trev}{\ensuremath{T_{i..j}}\xspace}
\newcommand{\Sfront}{\ensuremath{S_{f}}\xspace}
\newcommand{\Sback}{\ensuremath{S_{b}}\xspace}

\renewcommand{\_}{\raisebox{.6mm}{\text{\tiny\boldmath$-$}}}
\newcommand{\SSBR}{\textsc{Signed Sorting by Reversals}\xspace}

\newcommand{\alg}[1]{\textsc{#1}\xspace}
\newcommand{\algarg}[2]{\textsc{#1}(\ensuremath{#2})\xspace}

\newcommand*\Let[2]{\State #1 $\gets$ #2\xspace}
\renewcommand*\Return[1]{\State \textbf{return} #1}
\newcommand*\TRUE{\texttt{true}\xspace}
\newcommand*\FALSE{\texttt{false}\xspace}

\newtheorem{theorem}{Theorem}
\newtheorem{lemma}{Lemma}
\newtheorem{remark}{Remark}
\newtheorem{corollary}{Corollary} %[lemma]
\theoremstyle{definition}
\newtheorem*{definition}{Definition}

\newcommand{\mybox}[2]{
  \noindent\begin{tikzpicture}
    \node[minimum width=\linewidth-0.4pt, draw, rounded corners, text width=\linewidth-12pt] (a){#2};
    \node[fill=white, xshift=1em, anchor=west] at (a.north west) {#1};
  \end{tikzpicture}
}

\newcommand{\optproblem}[6]{
  \label{#6}
  \noindent\mybox{%
    \ifthenelse{\equal{#4}{}}{}{{\textsc{#4}}\ifthenelse{\equal{#5}{}}{}{ ({#5})}}
  }{%
    \begin{compactdesc}
      \item [Input:] {#1}
      \item [Output:] {#2}
      \item [Measure:] {#3}
    \end{compactdesc}
  }
}

\title{A Simple and Efficient Algorithm for\\
       Sorting Signed Permutations by Reversals}
\author{Krister M. Swenson}
\date{}

\begin{document}
\maketitle

\begin{abstract}
In 1937, biologists Sturtevant and Tan posed a computational question: transform a chromosome represented by a permutation of genes, into a second permutation, using a minimum-length sequence of \emph{reversals}, each inverting the order of a contiguous subset of elements.
Solutions to this problem, applied to \emph{Drosophila} chromosomes, were computed by hand.
The first algorithmic result was a heuristic that was published in 1982.
In the 1990s a more biologically relevant version of the problem, where the elements have signs that are also inverted by a reversal, finally received serious attention by the computer science community.
This effort eventually resulted in the first polynomial time algorithm for \alg{Signed Sorting by Reversals}.
Since then, a dozen more articles have been dedicated to simplifying the theory and developing algorithms with improved running times.
The current best algorithm, which runs in $O(n \log^2 n / \log\log n)$ time, fails to meet what some consider to be the likely lower bound of $O(n \log n)$.
In this article, we present the first algorithm that runs in $O(n \log n)$ time in the worst case.
The algorithm is fairly simple to implement, and the running time hides very low constants.
\end{abstract}

% % % % % % % % % % % % % % % % % % % % % % % % % % % % % % % % % % % % % % % %
% % % % % % % % % % % % % % % % % % % % % % % % % % % % % % % % % % % % % % % %

% - - - - - - - - - - - - - - - - - - - - - - - - - - - - - - - - - - - - - - -
\section{Introduction}

A century ago Alfred Sturtevant initiated the study of genome rearrangements \cite{sturtevantCaseRearrangementGenes1921}, eventually discovering and using ``inversions'' in \emph{Drosophila} chromosomes to reconstruct phylogenetic histories~\cite{sturtevantSequenceCorrespondingThirdchromosome1926,sturtevantInversionsThirdChromosome1936}.
A computational problem was posed by \citet{sturtevantComparativeGeneticsDrosophila1937} in 1937: transform one gene order represented by a permutation of elements $1,2,\dots,n$, into a second, using a minimum-length sequence of \emph{reversals} that each inverts the order of a contiguous set of elements.
Solutions were computed by hand~\cite{sturtevantHomologiesChromosomeElements1941}, which was an error-prone process that lead to confusion~\cite{tannierHaplessMathematicalContribution2022}.
The first algorithmic work on the subject was a heuristic by \citet{wattersonChromosomeInversionProblem1982} published in 1982.

%The problem remained invisible to the computer science community until a decade later when \citet{sankoffEditDistanceGenome1992} posed a more biologically relevant version of the problem, where each element possesses a sign representing the direction in which the gene is transcribed.
The computer science community remained unaware of the problem until a decade later when \citet{sankoffEditDistanceGenome1992} addressed a more biologically relevant variant, where each element possesses a sign representing the direction in which the corresponding gene is transcribed.
In this model, a reversal inverts both the order of contiguous elements, as well as their signs.
In the last 30 years this \emph{signed sorting by reversals} problem has received a significant amount of attention; we say ``sorting'' since any pair of permutations can be renamed to make one of them the identity permutation $(1~2~\cdots~n)$, and thus the problem can be stated in terms of a single input.

After initial explorations~\cite{sankoffEditDistanceGenome1992,kececiogluEfficientBoundsOriented1994}, efficient approximation algorithms were developed by \citet{kececiogluExactApproximationAlgorithms1995} and \citet{bafnaGenomeRearrangementsSorting1996}.
The first breakthrough was the famous result of \citet{hannenhalliTransformingCabbageTurnip1999}, which was an $O(n^4)$ algorithm based on the characterization and classification of hard-to-sort combinatorial structures in the permutations, relying on complicated transformations and an exhaustive search through all possible reversals at each step.

\bigskip
Hannenhalli and Pevzner call certain combinatorial structures of a permutation \emph{components}, and classify them into those that are \emph{oriented} and easier to sort, and into those that are \emph{unoriented} and are harder to sort.
In this article we choose the terminology of \citet{S.MIntroductionComputationalMolecular1997}, using \emph{good} and \emph{bad} in place of oriented and unoriented.
This classification into good and bad components is the first of the two essential tasks for sorting by reversals, while the second is the online maintenance of the components, whereby one can avoid choosing a reversal that creates a new bad component.
\bigskip

%There are two essential tasks for sorting by inversions: 1) the \emph{classification} of the components, and 2) the online \emph{maintenance} of this information, whereby one can avoid reversals that create new bad components.

Improving the classification into good and bad, was the focus of two studies.
First, \citet{bermanFastSortingReversal1996} achieved an $O(n\alpha(n))$ running time for the classification of components, leading to an $O(n^2\alpha(n))$ algorithm for \SSBR, where $\alpha(n)$ is the inverse Ackermann function.
Bader, Moret, and Yan~\cite{baderLinearTimeAlgorithmComputing2001} then met the linear-time lower bound for classification, yielding an optimal algorithm for computing the signed reversal distance, without calculating a sorting sequence.

All running time improvements for finding a sorting sequence required advances in the online update of the permutation and its components through each reversal.
A quadratic bound was reached by Kaplan, Shamir, and Tarjan~\cite{kaplanFasterSimplerAlgorithm2000}, based on a linear-time update.
Their key tool was the \emph{overlap graph} for a permutation, along with its implicit, and therefor efficient, representation.
A simplified presentation of the theory on the graph, and an algorithm based on bit-vector operations, was presented by \citet{bergeronVeryElementaryPresentation2005}, but this algorithm was shown to have running time limitations due to its explicit representation of the overlap graph~\cite{ozery-flatoTwoNotesGenome2003}.
For a more complete treatment of early developments in \SSBR, along with a historical perspective, we refer the interested reader to the chapter of \citet{bergeronInversionDistanceProblem2005}.

\citet{kaplanEfficientDataStructures2003} adopted a datastructure from the travelling salesman literature to finding a sorting sequence in subquadratic time, although there exist permutations on which they got ``stuck'' and could not provide a solution.
Tannier, Bergeron, and Sagot~\cite{tannierAdvancesSortingReversals2007} remedied this by developing a clever recovery scheme once stuck, yielding an $O(n \sqrt{n \log n})$ algorithm that works on any input permutation.
\citet{Han06:14} used B-Trees to improve the update of a permutation through reversal, from $O(\sqrt{n \log n})$ to $O(\sqrt{n})$.
\citet{rusuSortingSignedPermutations2018} introduced a datastructure for online update based on subset distinguishing sets, yielding a running time of $O(n(\frac{n}{b}\log b))$, where $b$ must be limited to the computer word size in order to ensure that the basic operations of the algorithm can be done in constant time.
We gave an algorithm that ran in $O(n \log n + kn)$ time, where $k$ is the number of times we got stuck~\cite{swensonSortingSignedPermutations2010}.
Recently Dudek, Gawrychowski, and Starikovskaya~\cite{D.G.SSortingSignedPermutations2024} devised an algorithm with running time $O(n \log^2n / \log\log n)$, based on the efficient maintenance of connectivity information on a newly conceived graph.

\bigskip
While these polynomial time results for \SSBR shine on the backdrop of the NP-hardness result of \citet{capraraSortingReversalsDifficult1997} for the unsigned version of the problem, the question remained open as to whether there exists an algorithm that runs in $O(n \log n)$ time.

In this article we provide an affirmative answer to the question, with an algorithm that is relatively easy to implement, while having low hidden constants.
We achieve the $O(n \log n)$ running time by re-adapting our binary search tree of \cite{swensonSortingSignedPermutations2010} to an efficient use of the recovery scheme devised by \citet{tannierAdvancesSortingReversals2007}.

%   -   -   -   -   -   -   -   -   -   -   -   -   -   -   -   -   -   -   -
\subsection{Basic definitions and notation}

\begin{figure}[]
  \begin{center}
    \includegraphics[alt={A simple example of a reversal sequence.},width=0.9\textwidth]{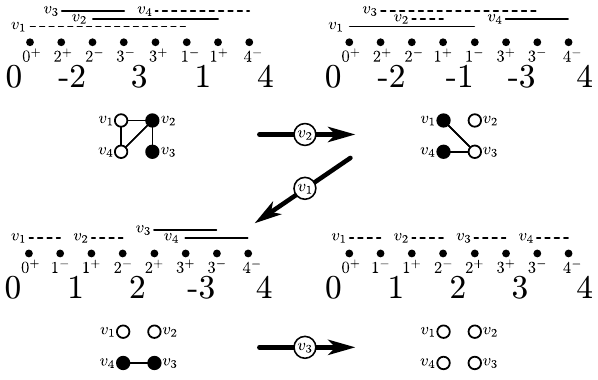}
  \end{center}
  \caption{A reversal sequence transforming $(0~\_2~3~1~4)$ into the identity permutation.
At the top, $(0~\_2~3~1~4)$ is transformed into $(0~\_2~\_1~\_3~4)$ by the first reversal $\rho(2,3) = \mu(\_2,1)$.
The pairs of points for the elements of a permutation appear in a line above the permutation, and an interval for each identity pair is indicated above those points by a black line;
lines are solid for good pairs and dashed for bad pairs.
The overlap graph for each permutation appears below it; good vertices are colored black and bad are colored white.
The sequence of reversals on permutations can be represented by the sequence of vertex complementations indicated by arrows, yielding a graph with only isolated white vertices.
}
\label{fig:example}
\end{figure}

Consider a permutation $(\pi_1~\pi_2~\cdots~\pi_n)$ of the set $\{1,2,\ldots,n\}$, where each element may be positive or negative.
To simplify the exposition we adopt the usual extension by adding $\pi_0 = 0$ and $\pi_{n+1} = n+1$ to the permutation, resulting in $\pi = (0~\pi_1~\pi_2~\cdots~\pi_n~n+1).$
%The negation of an element $\pi_i$ is denoted $\_\pi_i$, so that if for example $\pi_i = \_4$, then $\_\pi_i = 4$.
%, while the absolute value of an element $\pi_i$ is denoted by $|\pi_i|$.
A \emph{reversal} $\rho(i,j)$, for $1 \leq i < j \leq n$, on a permutation
$$\pi = (0~\pi_1~\cdots~\pi_i~\cdots~\pi_j~\cdots~\pi_n~n+1)$$
reverses all elements between $\pi_i$ and $\pi_j$ while changing their signs, yielding
$$\pi\rho(i,j) = (0~\pi_1~\cdots~\pi_{i-1}~\_\pi_j~\cdots~\_\pi_i~\pi_{j+1}~\cdots~\pi_n~n+1).$$
We denote the application of a sequence of reversals $S = \rho_1\rho_2\cdots\rho_m$ to $\pi$ as $\pi \cdot S = (\pi \cdot \rho_1 \cdots \rho_{m-1})\rho_m$, and the concatenation of two reversal sequences $S$ and $S'$ as $S \cdot S'$.
%The \emph{span} of a reversal $\rho(i,j)$ is the closed interval on the natural numbers $[i,j]$ and two spans $[i,j]$ and $[k,l]$ \emph{overlap} if and only if either $i < k$ and $k < j$ or $k < i$ and $j < l$.
The \SSBR problem calls for minimum-length sequence $S$ such that $\pi \cdot S$ is the identity.
Figure~\ref{fig:example} depicts such a sequence for $\pi = (0~\_2~3~1~4)$.

\bigskip
\optproblem%
{Signed permutation $\pi$.}%
%{Reversal sequence $S = \rho_1, \rho_2, \dots, \rho_d$ such that $\pi \cdot S$ is the identity permutation.}%
{Reversal sequence $S$ such that $\pi \cdot S$ is the identity permutation.}%
{Minimize length $|S|$ of the sequence.}%
{Signed Sorting by Reversals}{}{prob:SSR}

A pair of elements $(\pi_i, \pi_j)$, for $i,j \in \{0 \dots n+1\}$, is an \emph{identity pair} if $\big| |\pi_i| - |\pi_j| \big| = 1$, \ie, they appear consecutively in the identity permutation.
An identity pair is \emph{good} if $\pi_i$ and $\pi_j$ have opposite signs, otherwise it is \emph{bad}.
%Two consecutive elements $\pi_i$ and $\pi_{i+1}$ form an \emph{adjacency}.
%An adjacency is \emph{resolved} if $\pi_{i+1} - \pi_i = 1$, otherwise it is a \emph{breakpoint}.
Two consecutive elements $\pi_i$ and $\pi_{i+1}$ are called an \emph{adjacency} if $\pi_{i+1} - \pi_i = 1$, otherwise it is a \emph{breakpoint}.
In the example of Figure~\ref{fig:example}, the permutation in the top right has a single adjacency \adj{\_2}{\_1}, while all other consecutive elements form breakpoints.
The good pairs for this permutation are $(0, \_1)$ and $(\_3, 4)$.
For Figure~\ref{fig:bigexample}, the permutation $\pi'$ has adjacencies \adj{\_4}{\_3} and \adj{7}{8}.
The good pairs of $\pi'$ are $(\_5, 6)$ and $(0, \_1)$, while all other identity pairs are bad.

The reversal induced by a good pair $(\pi_i, \pi_j)$ is called a \emph{good} reversal, and is written as
\begin{equation*}
\mu(\pi_i, \pi_j) = 
\begin{cases}
  \rho(i, j-1) & \text{if } \pi_i + \pi_j = 1,\\
  \rho(i+1, j) & \text{if } \pi_i + \pi_j = \_1.
\end{cases}
\end{equation*}
A good reversal creates one or two adjacencies, two being created when two different good pairs imply the same reversal $\mu(\pi_i, \pi_{j-1}) = \mu(\pi_{i+1}, \pi_j) = \rho(i+1, j-1)$.
The good pair $(\_2, 1)$ in $(0~\_2~3~1~4)$ from Figure~\ref{fig:example} induces the reversal $\rho(3, 4)$.
The good pair $(3, \_4)$ in $\pi$ from Figure~\ref{fig:bigexample} induces the reversal $\rho(5, 10)$ that transforms $\pi$ into $\pi'$.
Note that a reversal will change some pairs from bad to good, and vice versa.
Thus, when we refer to a good reversal $\rho_i$ in a reversal sequence $\rho_1\rho_2\cdots\rho_m$ on $\pi$, it means that $\rho_i$ is a good reversal on permutation $\pi \cdot \rho_1\rho_2\cdots\rho_{i-1}$.
By extension, a \emph{good sequence} contains only good reversals.

A \emph{subpermutation} of $\pi$ is an ordered set of consecutive elements in $\pi$.
Roughly speaking a \emph{framed common interval} (FCI) of $\pi$ is a subpermutation that is framed by a smallest integer $a$ and largest integer $b$ while containing a (possibly empty) signed permutation $\pi_i~\pi_{i+1} \cdots \pi_j$ of all integers greater than $a$ and smaller than $b$~\cite{bergeronCommonIntervalsSorting2002}.
It is ``common'' with the identity permutation, since it also appears there.
Formally, an FCI is a subpermutation \mbox{$(a~\pi_i~\pi_{i+1}~\cdots~\pi_j~b)$} or \mbox{$(\_b~\pi_i~\pi_{i+1}~\cdots~\pi_j~\_a)$} of $\pi$, such that
\begin{enumerate}
  \item $a < b$,
  \item $(\pi_i~\pi_{i+1}~\cdots~\pi_j)$ is a (possibly empty) signed permutation of the integers $\{a+1, a+2, \dots, b-2, b-1\}$,
  \item it is not a union of shorter subpermutations with the first two properties.
\end{enumerate}
Integers $a$ and $b$ are the \emph{frame} elements and $\{\pi_i, \pi_{i+1}, \ldots, \pi_j\}$ are called the \emph{internal} elements of the FCI.
An FCI $B$ is \emph{nested} in an FCI $A$ if and only if the left and right frame elements of $A$ occur, respectively, before and after the frame elements of $B$.
The permutation $\pi$ in Figure~\ref{fig:bigexample} has a single FCI, whereas $\pi' = (0~\_5~\_2~\_4~\_3~\_1~6~10~9~7~8~11)$ has five FCIs, three of which are nested inside at least one other FCI.
For example, FCI $(\_4~\_3)$ is nested inside $(\_5~\_2~\_4~\_3~\_1)$, which is in turn nested in $(0~\_5~\_2~\_4~\_3~\_1~6)$.

A \emph{component} of $\pi$ is a set containing the frame elements of an FCI along with all internal elements of the FCI that are not an internal element of a nested FCI.
As components with less than four elements represent suites of adjacencies, we are interested in \emph{non-trivial components} with at least four elements.
A component is \emph{bad} if it is non-trivial, and all of its elements have the same sign, otherwise it is \emph{good}.
Note that two components can intersect only at their frame elements.
For example, $\pi'$ has components $\{\_4,\_3\}$, $\{\_5,\_2,\_4,\_3,\_1\}$, $\{0,\_5,\_1,6\}$, $\{7,8\}$, and $\{6, 10, 9, 7, 8, 11\}$.
Two of the components of $\pi'$ are trivial, one is good, and two are bad.

A reversal is called \emph{unsafe} if it creates a bad component, otherwise it is \emph{safe}.
%A permutation is \emph{positive} when every element in it is positive, indicating that either the permutation is the identity, or there exists at least one bad component.
Reversal $\rho(5, 10)$ in Figure~\ref{fig:bigexample} is an unsafe reversal on $\pi$, since it creates bad component $\{6, 10, 9, 7, 8, 11\}$, having elements that are all of the same sign.
Working with unsafe reversals is the main obstacle when sorting by reversals, as they must first be detected before being carefully transformed.
%The overview, later in this section will introduce this process in more detail.

%   -   -   -   -   -   -   -   -   -   -   -   -   -   -   -   -   -   -   -
\subsection{Overview}
\label{sec:overview}

The task is to construct a minimum length sequence of reversals that sorts a permutation $\pi$.
Hannenhalli and Pevzner proved that this can be reduced to the problem of constructing a good sequence of reversals.
\begin{theorem}[\citet{hannenhalliTransformingCabbageTurnip1999}]
\label{thm:hp}
A reversal sequence $S$ transforming a permutation into the identity is of minimum length if $S$ contains only good reversals.
\end{theorem}
\noindent Each good reversal corresponds to at least one good pair, and it turns out that it is convenient to build a sequence of good pairs, rather than a sequence of reversals.
Thus, we refer to good reversals and good pairs synonymously throughout the text.

We work in three domains.
Reversals on permutations are related to local complementations of vertex neighborhoods on a particular type of graph.
Certain proofs are easier to articulate in this realm, and Algorithm~\ref{alg:sort_graph} works on graphs but is difficult to make efficient.
In order to rapidly compute sequences of good pairs we must operate in the realm of balanced binary search trees.
The algorithm presented in Section~\ref{sec:algorithm} mimics Algorithm~\ref{alg:sort_graph}, while using the tree as the object to be transformed.
The results of Section~\ref{sec:recovery} are most easily stated directly on the permutations themselves.

We assume that $\pi$ does not contain any bad components, since any permutation can be transformed into one without a bad component in linear time, without effecting optimality.
Say that $\pi$ contains at least one bad component.
There exists a minimum-length sequence $S = S' \cdot S''$ where $\pi \cdot S$ is the identity permutation, $\pi \cdot S'$ has no bad components, and $S'$ can be calculated in linear time \cite{baderLinearTimeAlgorithmComputing2001,bergeronCommonIntervalsSorting2002}.

\bigskip
The general overview of the algorithm is that we continue building a sequence $S$ of reversals, until there is no good pair in $\pi \cdot S$.
At this point, if $\pi \cdot S$ is the identity permutation, then $S$ is of minimum length and we are done.
Otherwise, there must have been a bad component created by an unsafe reversal.

If we are stuck without a good pair, we recover by backtracking through reversals until just before the most recent bad reversal, at which point there must exist good pairs in the bad components created by that reversal.
With care, some of these good pairs can then be inserted into the sequence; the subsequent reversals that we undid in the backtracking will change, but their corresponding good pairs will not.
Section~\ref{sec:tannier} revisits this framework of \citet{tannierAdvancesSortingReversals2007} by explicitly detailing lemmas that they implicitly used, while adding lemmas that interface more easily with our results.
In this section it is most convenient to prove lemmas about vertex complementations on graphs.

Section~\ref{sec:datastructure} presents an evolution of the previously used balanced binary search tree from \cite{swensonSortingSignedPermutations2010}, adapted to the recovery scheme.
Section~\ref{sec:recovery} proves that the datastructure can perform the backtracking and bad reversal detection efficiently.

While Section~\ref{sec:tannier} presents a new recursive version of the \citet{tannierAdvancesSortingReversals2007} algorithm along with a proof of correctness, it is not obvious how to directly make this algorithm efficient.
Section~\ref{sec:algorithm} presents the efficient version of the same algorithm, which operates on the search tree of Section~\ref{sec:datastructure}, rather than on permutations or on graphs.

In the rest of the section, we finish developing the basic definitions and notions that will be used throughout rest of the article.

%   .   .   .   .   .   .   .   .   .   .   .   .   .   .   .   .   .   .   .
\subsection{A note on shared notation}

In the following section we will define the \emph{overlap graph} for a permutation, relating \emph{vertex complementations} on the graph to reversals on permutations.
When concepts from the two domains relate to each other, we will share terminology, and sometimes notation.
Good and bad pairs in permutations correspond to good (black) and bad (white) vertices in the overlap graph, good and bad components in the permutation correspond to good and bad components in the overlap graph, and good reversals correspond to local complementations of good vertices.
An unsafe reversal is, then, a complementation of a good vertex that creates a new all-white component.

\begin{figure}[t]
  \begin{center}
    \includegraphics[alt={An unsafe reversal.},width=1.0\textwidth]{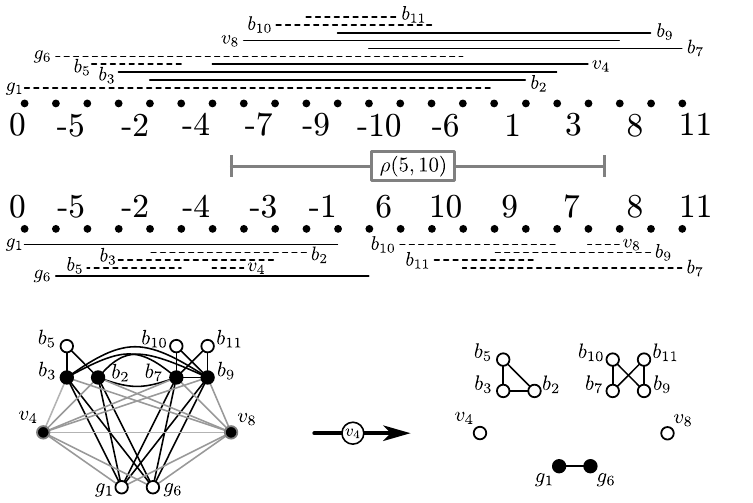}
  \end{center}
  \caption{
The upper permutation $\pi = (0~\_5~\_2~\_4~\_7~\_9~\_10~\_6~1~3~8~11)$ is transformed into $\pi' = \pi\rho(5, 10) = (0~\_5~\_2~\_4~\_3~\_1~6~10~9~7~8~11)$ through the unsafe reversal $\rho(5, 10) = \mu(3, \_4) = \mu(\_7, 8)$.
The overlap graph for $\pi$ is on the bottom-left, and the overlap graph for $\pi'$ is on the bottom-right.
Vertices from the bad components of $OV(\pi')$ are labeled by $b$ and vertices from the good components are labeled by $g$.
See the caption of Figure~\ref{fig:example} for a more detailed description.
}
\label{fig:bigexample}
\end{figure}

The notion of ``restriction'' will be used for graphs and for permutations alike.
In the overlap graph, a \emph{restriction} is simply an induced subgraph; for a set of vertices $W \subseteq V$ of a graph $H = (V, E)$, the restriction $H[W]$ of $H$ to $W$ is the graph on vertices $W$, having exactly the edges in $E$ with both endpoints in $W$.
For a permutation $\pi$ and a set of elements $Q \subseteq \{1,\dots,n\}$, the \emph{restriction} $\pi[Q]$ of $\pi$ to $Q$ is the permutation $\pi$ with elements not in $Q$ removed.
For $\pi$ in Figure~\ref{fig:bigexample} and $Q = \{6,7,8,9,10,11\}$, we have $\pi[Q] = (6~10~9~7~8~11)$.

%   -   -   -   -   -   -   -   -   -   -   -   -   -   -   -   -   -   -   -
\subsection{The overlap graph}
\label{sec:overlapgraph}

While in Section~\ref{sec:recovery} we work directly on permutations, in the next section the results are most easily articulated in the domain of graphs.
%While the results of Section~\ref{sec:complementation} are applicable to any graph having vertices that are labeled as \emph{good} or \emph{bad}, the results of Section~\ref{sec:sortingOG} pertain to a more specific graph called an overlap graph.
\citet{kaplanFasterSimplerAlgorithm2000} associated to a permutation $\pi$ a specific type of circle graph~\cite{GavrilAlgorithmsMaximumClique1973}, called an overlap graph.
Roughly speaking, an overlap graph is an intersection graph of the intervals defined by the identity pairs.

Given a permutation $\pi = (0~\pi_1~\cdots~\pi_n~n+1)$, we embed a set of $2n+2$ points on a line, ordered according to the permutation.
There is a pair of adjacent points $\pi_i^-$ and $\pi_i^+$ for each element $\pi_i$, and the points for $\pi_i$ come before the points for $\pi_j$ if and only if $i < j$.
Point $\pi_i^-$ appears before $\pi_i^+$ if $\pi_i$ is positive, but after it if $\pi_i$ is negative.
For element $0$ there is the first point $0^+$ on the line, and for $n+1$ there is the last point $(n+1)^-$.
%while the pairs of points for the $\pi_i$'s are ordered between $0^+$ and $(n+1)^-$ according to the permutation.
Each pair of points $(q^+, (q+1)^-)$ corresponds to an interval on the line.
In Figure~\ref{fig:example} the points $\{0^+, 1^-, 1^+, 2^-, 2^+, 3^-, 3^+, 4^-\}$ are aligned above each of the four genomes.
Above the points we have a line for each of the $4$ intervals $(q^+, (q+1)^-)$, for $0 \leq q < 4$.
%Each pair of points $(q^+, (q+1)^-)$, for $0 \leq q \leq n$, corresponds to an interval on the line.

The \emph{overlap graph} $OV(\pi)$ for permutation $\pi$ has a vertex for each of the intervals $(q^+, (q+1)^-)$, for $0 \leq q \leq n$.
There is an edge $(u, v)$ if the interval $u$ overlaps with interval $v$, without one interval being contained within the other.
A vertex $(q^+, (q+1)^-)$ is \emph{good} if the pair $(q, q+1)$ is good, otherwise it is \emph{bad}.
Below each permutation in Figure~\ref{fig:example} appears its overlap graph.

The overlap graph construction implicates a bijection between the components of a permutation $\pi$ and the connected components of $OV(\pi)$~\cite{bergeronCommonIntervalsSorting2002}.
Reversals on permutations, then, correspond to an operation called a local complementation of a vertex in the graph.

The local complementation operation is defined on a graph $H = (V, E)$, where each vertex is colored either black or white, corresponding in our case to a vertex that is good or bad, respectively.
The \emph{neighborhood} of a vertex $v \in V$, denoted $N(v)$, is the set of vertices adjacent to $v$ in $H$.

\begin{definition}
The \emph{local complementation} of a good vertex $v \in V$, denoted $H/v$, complements the neighborhood $W = N(v) \cup \{v\}$, while toggling all vertices in $W$ from good to bad, or vice versa.
That is, there is an edge $e$ in $H/v$ if and only if $e$ is not in $H[W]$ or $e$ is in $H[V \setminus W]$.
\end{definition}

\noindent Note, then, that $v$ is always isolated and bad in $H/v$.

%For a sequence $S = v_1v_2 \cdots v_m$ of vertices, define $H/S = H/v_1v_2 \cdots v_m$ as $(H/v_1 \cdots v_{m-1})/v_m$.
The task of sorting by reversal can be reduced to that of ``sorting'' a graph, by finding a sequence of complementations of good vertices $v_1v_2 \cdots v_m$ that yields only isolated vertices in $H/v_1v_2 \cdots v_m$.

% - - - - - - - - - - - - - - - - - - - - - - - - - - - - - - - - - - - - - - -
\section{Tannier, Bergeron, and Sagot's approach}
\label{sec:tannier}

The main contribution of \citet{tannierAdvancesSortingReversals2007} was a clever technique for recovering from an unsafe reversal.
They construct a sequence of reversals $S = \rho_1\rho_2 \ldots \rho_m$ by always choosing a good reversal as long as one exists, knowing that the absence of such a reversal implies that either the permutation is sorted, or that an unsafe reversal exists somewhere in $S$.
The key observation they make is that all of the reversals after the most recent unsafe reversal $\rho_i$ can be applied at the end of the sorting sequence, when the proper precautions are made.
To that end, they presented the following theorem.
\begin{theorem}[\citet{tannierAdvancesSortingReversals2007}]
\label{thm:split}
For any sequence $S$ of good reversals such that $\pi \cdot S$ has all positive elements, there exists a nonempty reversal sequence $S'$ such that $S$ can be split into two $S = S_1 \cdot S_2$, and $S_1 \cdot S' \cdot S_2$ is a good sequence on $\pi$.
\end{theorem}

While the algorithm that they developed is inspired by this theorem, some of the elements of the algorithm's correctness are hidden in the proof of the theorem, and some are left implicit, making it difficult for us to use their results directly.
Therefore, in this section we motivate a new recursive variant of the algorithm, while providing a thorough proof of correctness.
The general structure of these results is then used in our algorithm presented in Section~\ref{sec:algorithm}.

%   .   .   .   .   .   .   .   .   .   .   .   .   .   .   .   .   .   .   .
\subsection{Sorting by complementation}
\label{sec:complementation}

In \cite{tannierAdvancesSortingReversals2007}, Theorem~\ref{thm:split} is actually stated in a more general form that applies to local complementations on a bi-colored graph.
Their approach is based on the observation that an overlap graph has a particular structure, just before a bad component is created by an unsafe complementation of a vertex $v$.
Due to this structure, the complementation of $v$ can be made safe by inserting vertices before it in the already computed complementation sequence.

We first outline the structure in two remarks, before exploring the consequences.
The general idea is to focus on the state of a graph just before an unsafe complementation.
In this section, we will call a graph in this state $H$, and the unsafe complementation $v$.
The vertices of $H$ can be partitioned into the set $B$ containing the bad components of $H/v$, and the set $G$ containing the good components.
We show that the sequence of vertices that sorts the induced subgraph $H[B]$ can always be done before the sequence of vertices that sorts $H[G]$.
This allows us to iteratively proceed until we get stuck, having no good vertices to act on, before ``recovering'' by undoing the sequence on $H[G]$ and inserting a new sequence on $H[B]$ before $v$.

\begin{figure}[]
  \begin{center}
    \includegraphics[alt={Creation of bad components.},width=1.0\textwidth]{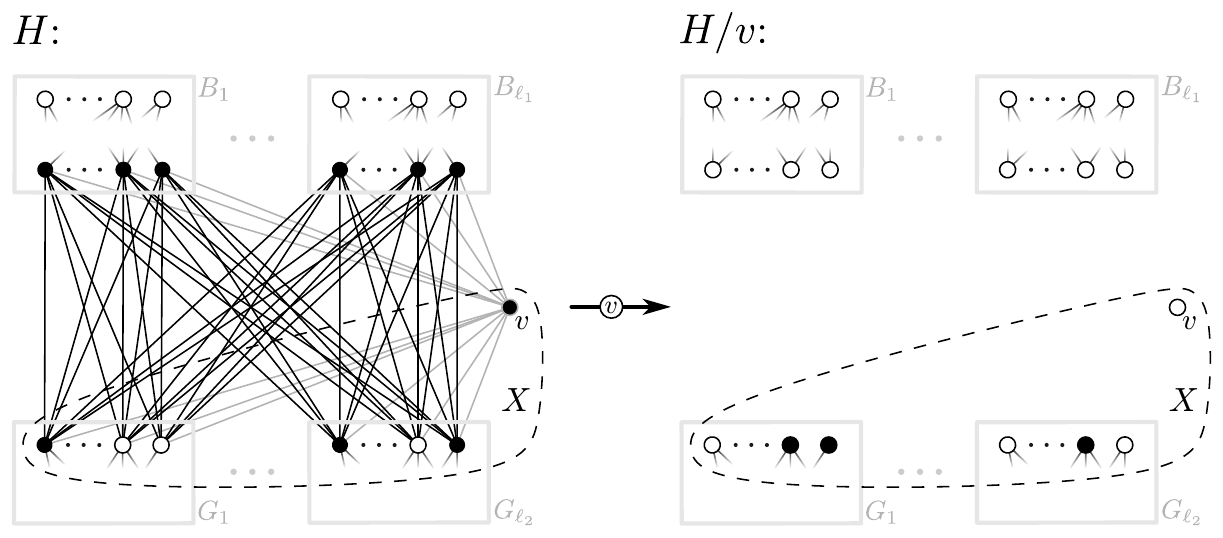}
  \end{center}
  \caption{
On the left, a graph $H$ has a single connected component, before being split into graph $H/v$ having bad components $B_1, B_2, \ldots, B_{\ell_{1}}$ and good components $G_1, G_2, \ldots, G_{\ell_{2}}$, on the right.
Good vertices are filled in black.
Note that, due to the definition of vertex complementation, all vertices from $B = \cup_{i=1}^{\ell_{1}} B_i$ that are adjacent to $v$ must be good, while the others from $B$ must be bad.
If, instead of $v$, a good vertex $u$ from $B$ is applied to $H$, then any good vertex in $H/u[B]$ will be adjacent to all the vertices of $X$.
%If, instead of $v$, a sequence of good vertices from $B$ are applied to $H$, then each of these vertices will be adjacent to all the vertices of $X$ at the time of its complementation.
}
\label{fig:components}
\end{figure}

The following remark is a simple consequence of the definition of local vertex complementation.
See Figure~\ref{fig:components} for an illustration.
\begin{remark}\label{rem:complementation}
The complementation of a vertex $v$ will split a single component into $\ell$ new components $C_1, C_2, \ldots, C_\ell$ if and only if for all $1 \leq i < j \leq \ell$
\begin{enumerate}
\item\label{item:alledges}
there are all possible edges between $C_i \cap N(v)$ and $C_j \cap N(v)$, and
\item\label{item:noedges}
there is no edge between $C_i \setminus N(v)$ and $C_j \setminus N(v)$.
\end{enumerate}
\end{remark}

Consider a graph $H$ with a single connected component, where the complementation of vertex $v$ creates the graph $H/v$ with bad components $B_1, B_2, \ldots, B_{\ell_{1}}$ and good components $G_1, G_2, \ldots, G_{\ell_{2}}$.
Call $B = \cup_{i=1}^{\ell_{1}} B_i$ the set of vertices that are in bad components of $H/v$, and $G = \cup_{i=1}^{\ell_{2}} G_i$ the set of vertices that are in good components of $H/v$.
The properties of Remark~\ref{rem:complementation} impose strong constraints on the colors of vertices in $B$, just before the unsafe complementation of vertex $v$ in $H$ (see Figure~\ref{fig:components}).
\begin{remark}\label{rem:goodinbad}
%If complementing vertex $v$ in graph $H$ creates a new bad component $C$ in $H/v$, then a vertex is bad in $H[C]$ if and only if it is in the neighborhood $C \cap N(v)$ of $v$ in $H$.
%If complementing vertex $v$ in graph $H$ creates a new bad component $B$ in $H/v$,
All vertices from $B$ that are adjacent to $v$ in $H$ are good, while the other vertices from $B$ are bad.
\end{remark}

The following lemma guarantees that, if instead of complementing $v$, we choose a vertex $u$ from $B$ to complement, we are guaranteed to have additional good vertices in $B$.
This is useful since it means we can build a complementation sequence of at least two good vertices, when recovering from the unsafe complementation of $v$.
\begin{lemma}\label{lem:anygood}
For any good vertex $u \in B$ in $H$, there exists at least one good vertex in $H/u[B]$.
\end{lemma}
\begin{proof}
If $u$ is adjacent to a bad vertex in $H[B]$, then the lemma is clearly true.
If $u$ is not adjacent to a bad vertex in $H[B]$, then there must exist some other vertex $u' \in B$ that is adjacent to $u$ in $H/v[B]$, otherwise $u$ would be isolated in $H/v[B]$, contradicting its membership in $B$.
Vertex $u'$, then, is not adjacent to $u$ in $H$ and is good in $H/u$.
\end{proof}

The following lemmas delineate the pertinent properties of the vertices that are adjacent to every member of $B \cap N(v)$ in $H$.
Call these vertices $X = (G \cap N(v)) \cup \{v\}$, as indicated in Figure~\ref{fig:components}.
The objective is to eventually show that any vertex sequence sorting the components of $B$ can be placed before any vertex sequence sorting the components of $G$.

\begin{lemma} \label{lem:Xconnectivity}
Consider a good vertex sequence $v_1v_2 \cdots v_m$ on $H$, composed from a subset of vertices in $B$.
Any good vertex in $H/v_1 \cdots v_m[B]$, has each of the vertices of $X$ in its neighborhood, and none of the vertices of $G \setminus X$.
\end{lemma}
\begin{proof}
Remarks~\ref{rem:complementation} and \ref{rem:goodinbad} demonstrate that the statement is true in $H$ (when $m = 0$).
%The statement is also clearly true for $v$ in $H$, since the complementation of $v$ would make all vertices of $B$ bad.
Say the property holds for $H/v_1 \cdots v_{i-1}$.
Then when we complement good vertex $v_i$, any good neighbor of $v_i$ will become disconnected from the vertices of $X$ and bad, while any bad neighbor of $v_i$ will become connected to $X$ and good.
\end{proof}

The following is a corollary of Lemma~\ref{lem:Xconnectivity} and Remark~\ref{rem:complementation}.\ref{item:noedges}, as each complementation in $v_1v_2 \cdots v_m$ will complement connectivity in $X$ while leaving the rest of $G$ unchanged.

\begin{corollary} \label{cor:even}
Consider a good vertex sequence $v_1v_2 \cdots v_m$ on $H$ of even length, composed from a subset of vertices in $B$.
Then $H/v_1 \cdots v_m[G \cup \{v\}] = H[G \cup \{v\}]$.
\end{corollary}

%\begin{lemma}\label{lem:vafter}
%Consider a good vertex sequence $v_1/v_2/\cdots/v_m$, composed from a subset of vertices in $B$, such that $H/v_1/\cdots/v_m[B]$ has no good vertex.
%In $H/v_1/\cdots/v_m$, vertex $v$ is either good or isolated.
%\end{lemma}
\begin{lemma}\label{lem:preserveG}
Consider a good vertex sequence $v_1v_2 \cdots v_m$, composed from a subset of vertices in $B$, such that $H/v_1 \cdots v_m[B]$ has no good vertex.
If $m$ is even then $H/v[G \cup \{v\}] = H/v_1 \cdots v_mv[G \cup \{v\}]$, otherwise $H/v[G \cup \{v\}] = H/v_1 \cdots v_{m}[G \cup \{v\}]$ if $m$ is odd.
\end{lemma}
\begin{proof}
When the length $m$ of the sequence is even, Corollary~\ref{cor:even} implies that $v$ is good in $H/v_1 \cdots v_m[G \cup \{v\}]$, and that $H/v[G \cup \{v\}] = H/v_1 \cdots v_mv[G \cup \{v\}]$.

If $m$ is odd, then $H[G \cup \{v\}] = H/v_1 \cdots v_{m-1}[G \cup \{v\}]$ by Corollary~\ref{cor:even}.
Since $v_m$ is good in $H/v_1 \cdots v_{m-1}$, Lemma~\ref{lem:Xconnectivity} implies that $v_m$ has exactly the vertices of $X$ in its neighborhood in $H/v_1 \cdots v_{m-1}$.
By the definition of $v$, it is adjacent to vertices $\{v_m\} \cup X \setminus \{v\}$ in $H[G]$.
Thus, the same adjacencies are complemented by $v$ in $H[G \cup \{v\}]$ as are complemented by $v_m$ in $H/v_1 \cdots v_{m-1}[G \cup \{v\}]$, and we have $H/v[G \cup \{v\}] = H/v_1 \cdots v_m[G \cup \{v\}]$.
\end{proof}

%   .   .   .   .   .   .   .   .   .   .   .   .   .   .   .   .   .   .   .
\subsection{Sorting overlap graphs}
\label{sec:sortingOG}

Lemma~\ref{lem:preserveG} inspires the following recursive algorithm for sorting an overlap graph with good components.
\begin{algorithm}[H]
\caption{Sort a good component by vertex complementation.}
\label{alg:sort_graph}
\begin{algorithmic}[0]   %1 for line numbers

\Statex $\triangleright$ Compute sorting sequence for good components of $H[Q]$.
\Procedure{SortGraph}{$H, Q$}
  \Let{$\Sfront, Q$}{\algarg{DoGood}{H, Q}}
  \State \Sback is an empty sequence
  \Comment{$f$ for ``front'' and $b$ for ``back''}
  \While{\Sfront is not empty}
    \Let{$S_1, S_2$}{\algarg{Recover}{H, \Sfront, Q}}
      \Comment{$S_1 \cdot S_2 = \Sfront$}
    \Let{$S', Q$}{\algarg{SortGraph}{H/S_1, Q}}
    \If{$even(|S'|)$}
       \Let{\Sback}{$S' \cdot S_2 \cdot \Sback$}
    \Else
       \Let{\Sback}{$S' \cdot S_2^- \cdot \Sback$}
       \Comment{$S_2^-$ is $S_2$ without the first vertex}
    \EndIf
    \Let{\Sfront}{$S_1$}
  \EndWhile
  \Return{$\Sback, Q$}
\EndProcedure

\Statex
\Statex $\triangleright$ Complement good vertices in $H[Q]$ until there are none.
\Procedure{DoGood}{$H, Q$}
  %\Let{$S$}{()}
  \State $S$ is an empty sequence
  \While{$H/S[Q]$ has a good vertex}
    \Let{$v$}{\algarg{GetGood}{H/S[Q]}} \Comment{any good vertex in $H/S[Q]$}
    %\State $append(S, v)$
    \Let{$Q$}{$Q \setminus \algarg{Isolated}{H/Sv[Q]}$} \Comment{remove newly isolated vertices}
    \Let{$S$}{$Sv$}
  \EndWhile
  \Return{$S, Q$}
\EndProcedure

\Statex
\Statex $\triangleright$ Get largest prefix $S_1$ of $S$ such that $H/S_1[Q]$ has a good vertex.
\Procedure{Recover}{$H, S, Q$}
  \State $S_2$ is an empty sequence \Comment{$S_1 \cdot S_2 = S$}
  \Let{$S_1$}{S}
  \While{$S_1$ is nonempty and $H/S_1[Q]$ has no good vertex}
    \Let{$v$}{$pop(S_1)$} \Comment{$pop()$ removes $v$ from the end}
    \Let{$S_2$}{$v \cdot S_2$}
  \EndWhile
  \Return{$S_1, S_2$}
\EndProcedure
\end{algorithmic}
\end{algorithm}

In the following, we show that Algorithm~\ref{alg:sort_graph} builds a sequence
of good vertices, which is guaranteed to be of minimum length due to
Theorem~\ref{thm:hp}.
%the following theorem, where $d(H)$ denotes the minimum vertex complementation
%distance for graph $H$.
%\begin{theorem}[\citet{hannenhalliTransformingCabbageTurnip1999}]
%If $v$ is a safe vertex in overlap an graph $H$, then $d(H/v) = d(H) - 1$.
%\end{theorem}\todo{refer to first theorem}
%Let us address the correctness of Algorithm~\ref{alg:sort_graph}.

Focus first on the $H$ and $Q$ within \alg{DoGood}.
Clearly, \alg{DoGood} returns a sequence $S$ of good vertices on $H[Q]$, and the $Q$ that is returned contains only vertices that are bad and unisolated in $H/S[Q]$.

Now focus on the $H$, $S$, and $Q$ within \alg{Recover}.
Clearly, \alg{Recover} returns the $S_1$ and $S_2$ such that $S = S_1 \cdot S_2$, the graph $H/S_1[Q]$ has a good vertex, and $S_2$ is of minimum length.

We now argue that a call to \algarg{SortGraph}{H, Q} computes a sorting sequence for $H[Q]$.
Sequence \Sfront is first build from good vertices in $H$.
The \textbf{while} loop iteratively backtracks through \Sfront using calls to \alg{Recover}; suffixes of \Sfront are moved to \Sback.

If $S_1$ is not empty after a call to \alg{Recover}, this indicates that there were bad components $B_1, B_2, \dots, B_{\ell_1}$ in $H/\Sfront[Q]$.
According to Remark~\ref{rem:goodinbad}, the graph $H/S_1[Q]$ has good vertices, which come from the bad components that we will call $B_1, \dots, B_k$, for $1 \leq k \leq \ell_1$.
%Note that $S_2$ is the sequence that sorts the good connected components created by the unsafe reversal on $H/S_1$.
Lemma~\ref{lem:preserveG} ensures that, after the recursive call to \algarg{SortGraph}{H/S_1, Q}, the sequence $S'$ that sorts components $B_1, \dots, B_k$ can safely be inserted between $S_1$ and $S_2$.

At the end of an iteration of the \textbf{while} loop, there are no good vertices;
this is obviously true when $|S'|$ is even, but is also true when $|S'|$ is odd due to similar reasoning as the proof of Lemma~\ref{lem:preserveG} (\ie complementation of the final vertex of $S'$ has the same effect on the graph as complementation of the first vertex of $S_2$).
The next iteration backtracks to find another vertex complementation that created some bad components,
and when \Sfront is empty, $Q$ is also empty and the graph $H$ has no remaining nontrivial components.
Each call to \alg{SortGraph} is independent from the previous, since each of them works on a disjoint subset of the bad components $B_1, \dots, B_{\ell_1}$.

The recursive call to \alg{SortGraph} is performed at most $n$ times, since by Lemma~\ref{lem:anygood} it will return a sequence with at least two vertices, and since each of the calls to \alg{SortGraph} works on an independent set of good components from the previous calls.
The running time of the algorithm, therefore, depends on the cost of the \alg{GetGood} call, the cost of doing or undoing a vertex complementation on the graph, and the cost of removing a vertex from $Q$; it is clear that each of these are done at most once for each vertex in the graph.

In the following sections we present a datastructure, along with the other necessary formalities, showing that each these steps can be performed in $O(\log n)$ amortized time.

% - - - - - - - - - - - - - - - - - - - - - - - - - - - - - - - - - - - - - - -
\section{Maximum and minimum pairs}
\label{sec:datastructure}

Previous approaches maintained datastructures containing the complete complement of good pairs which, as yet, has made it difficult to obtain an $O(n \log n)$ running time.
In \cite{swensonSortingSignedPermutations2010}, however, we ensured that we could always efficiently retrieve two specific good pairs of $\pi$: the pair that contains the maximum negative integer $\maxneg(\pi)$, and the pair that contains the minimum negative integer $\minneg(\pi)$.
In Figure~\ref{fig:bigexample} we have $\maxneg(\pi) = 2$, $\minneg(\pi) = 10$, $\maxneg(\pi') = 1$, and $\minneg(\pi') = 5$.

Since sorting by reversals is an iterative process, it is natural to speak in reference to an implied ``current'' permutation.
When it is clear from the context, we will drop the $\pi$ and just speak of the (``maximum negative'') \emph{\maxneg element} in the current permutation, along with its \emph{\maxneg pair}.

%The following lemma shows our motivation for maintaining this information.
%While it addresses the \maxneg pair, the \minneg pair works similarly.
%\begin{lemma}
%Consider a sequence $S$ of \maxneg reversals applied to $\pi$, and the set $N$ of negative elements from the good pairs used in $S$.
%The maximum negative element $q = \max(\n(\pi') \setminus N)$ of $\pi' = \pi \cdot S$, is the negative element of a good pair $(q-1, \_q)$ in $\pi'$.
%\end{lemma}
%\begin{proof}
%Call $r$ the smallest element $0 \leq r < q$ that is positive in $\pi'$, such that element $r+1$ is negative in $\pi'$.
%The element $r+1$ cannot be in $N$, since this would imply that the breakpoint between $r$ and $r+1$ has been healed by $S$, and that they have the same sign.
%Therefore, by definition, $q = r+1$.
%\end{proof}

The use of \maxneg and \minneg pairs in \cite{swensonSortingSignedPermutations2010} enabled us to find a good reversal in logarithmic time, yet had the drawback of requiring convoluted and inefficient recovery steps when no good reversal remained.
In this section we supplement the datastructure with subsets of the unsigned integers in $\pi$ that are still ``eligible'' to be a maximum/minimum negative element in a good pair, ensuring that, when recovering we can ignore pairs that have already been repaired by a reversal.
Call \Qmax the set containing every element $q \in \{1, \dots, n+1\}$ such that the adjacency \adj{(q-1)}{q} does not exist in the current permutation, and call \Qmin the set of elements containing $q \in \{0, \dots, n\}$, such that the adjacency \adj{q}{(q+1)} does not exist.
%The set of unsigned integers that are eligible to be the negative element for a \maxneg pair is called $\Qmax \subseteq \{1, \dots, n\}$, while the eligible set for the negative element in a \minneg pair is called $\Qmin \subseteq \{1, \dots, n\}$.
Thus, any possible good pair $(\pi_i, \pi_j)$ that does not yet form an adjacency, is represented by one element of \Qmax and one element of \Qmin.
For example when $|\pi_i| < |\pi_j|$ we have $|\pi_i| \in \Qmin$ and $|\pi_j| \in \Qmax$.

Maintaining these sets allows us to use simpler recovery steps within the framework of a modified version of the \citet{tannierAdvancesSortingReversals2007} algorithm; we prove in the next section that during recovery either the \maxneg element must be in \Qmax, or the \minneg element must be in \Qmin.
To simplify the exposition we use a splay tree, but other balanced binary search trees based on rotations (\eg AVL trees and Red-black trees) could be used to similar effect, thereby avoiding amortized running times.

%   -   -   -   -   -   -   -   -   -   -   -   -   -   -   -   -   -   -   -
\subsection{The datastructure}

Consider an ordered binary tree where each node $v$ represents an element $\pi_i$ of a signed permutation $\pi$,
noting that throughout this section we will refer to $v$ and $\pi_i$ interchangeably.
The initial conditions of the tree ensure that an in-order traversal visits vertices in the order that the elements appear in the permutation.
Thus, each subtree represents a subset of contiguous elements of $\pi$.

Along with subtree size, we associate to each node several values specific to our application.
The \emph{extremal values} for the subtree rooted at $v$ are the
\begin{itemize}
%\item $position(v)$ integer that contains the position of $v$ in $\pi$, a

\item $\maxneg(v)$ and $\maxpos(v)$ integers equal to the maximum negative/positive elements from \Qmax that appear in the subtree rooted at $v$, set to $\_\infty$/zero if none exist, and the

\item $\minneg(v)$ and $\minpos(v)$ integers equal to the minimum negative/positive elements from \Qmin that appear in the subtree rooted at $v$, set to zero/$\infty$ if none exist.
\end{itemize}
A $\rev(v)$ boolean flag indicates the subtree rooted at $v$ is in reverse order.
The $Q$ membership values for a vertex $v$ are the
\begin{itemize}
\item $\inQmax(v)$ boolean flag to indicate the element at $v$ is in \Qmax, and the

\item $\inQmin(v)$ boolean flag to indicate the element at $v$ is in \Qmin.
\end{itemize}

Note that a node $v$ has an implicit \emph{parity} based on the number of \rev flags on the path from $v$ to the root.
Say that the left child of $v$ is the \emph{first child} of $v$, and that the right child is the \emph{last child}, if the parity of $v$ is even.
Otherwise, the right child is the first, and the left child is the last.
An \emph{in-order} traversal of a tree, then, recursively visits before $v$ all nodes in the subtree rooted at the first child, and visits after $v$ all nodes in the subtree rooted at the last child.
Before any reversals have been done, the parity of every vertex is even since all \rev flags are initialized to \FALSE.

Similar to the in-order traversal we just described, the \rev flags act on the extremal values.
Since reversals flip the signs of the elements that are reversed, the maximum negative element from \Qmax in the subtree rooted at $v$ is $\maxneg(v)$ if the parity of $v$ is even, and $\maxpos(v)$ otherwise.

\begin{figure}[]
  \centering
  \begin{subfigure}{1.0\textwidth}
    \includegraphics[alt={A splay tree before reversal.},width=1.0\textwidth]{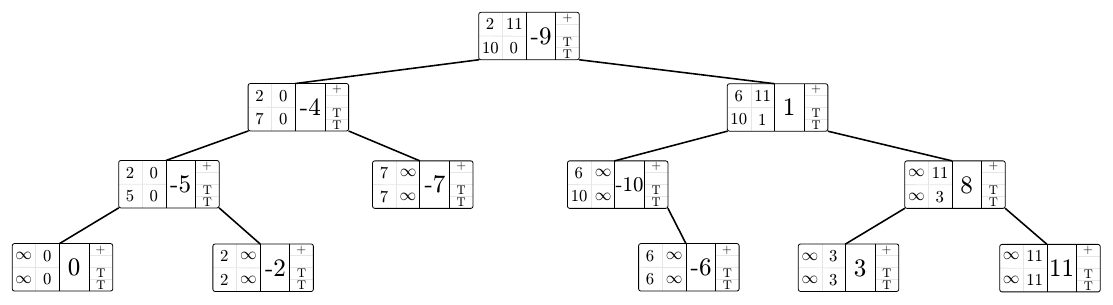}
    \caption{A tree from $\T(\pi, Q)$ where $\pi = (0~\_5~\_2~\_4~\_7~\_9~\_10~\_6~1~3~8~11)$, and $Q = (\{0, \dots, 11\}, \{0, \dots, 11\})$.}
    \label{fig:splaytree}
  \end{subfigure}
  \quad
  \begin{subfigure}{1.0\textwidth}
    \includegraphics[alt={A splay tree after reversal.},width=1.0\textwidth]{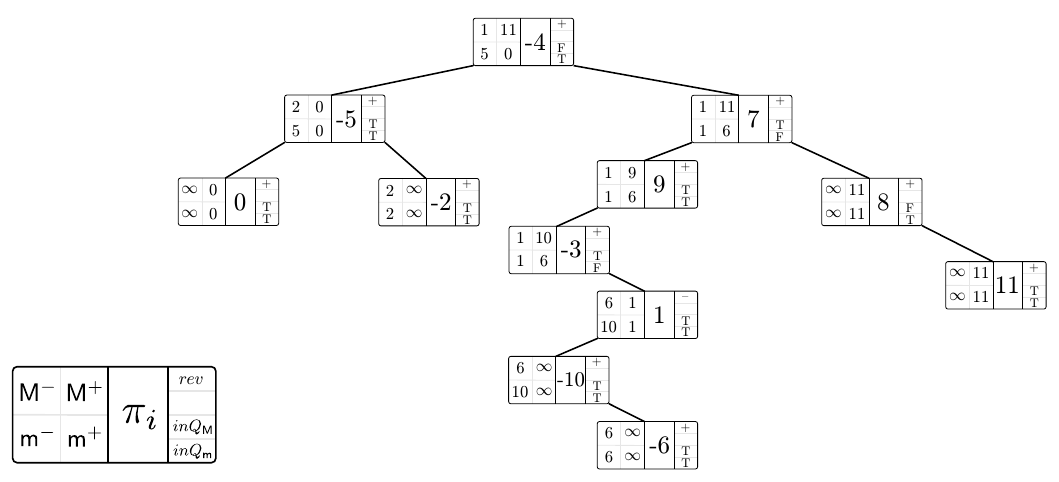}
    \caption{A tree from $\T(\pi', Q')$ where $\pi' = \pi\mu(\_7, 8) = (0~\_5~\_2~\_4~\_3~\_1~6~10~9~7~8~11)$,
    and $Q' = (\{0, 1, 2, 3, 5, 6, 7, 9, 10, 11\}, \{0, 1, 2, 4, 5, 6, 8, 9, 10, 11\})$.
	Note that the \rev flag in the node for element $1$ is set to true, indicated by the `$-$' in its corner.}
    \label{fig:reversedtree}
  \end{subfigure}
  \caption{A splay tree a) before, and b) after the reversal $\mu(\_7, 8)$ of Figure~\ref{fig:bigexample}.
Nodes are labeled by the values indicated in the bottom-left corner of the figure.
}%
\label{fig:reversal}
\end{figure}

\begin{definition}%[$\T(\pi, Q)$]
Call $\T(\pi, Q)$ the set of ordered trees, for a signed permutation $\pi$ and integer subsets $(\Qmax, \Qmin) = Q$, such that any tree $T \in \T(\pi, Q)$ has the following properties:
\begin{enumerate}
\item
$\pi$ can be retrieved by an in-order traversal of $T$, and

\item
extremal values $\maxneg(v)$ and $\minneg(v)$ can be retrieved by traversing $T$ from the root to any node $v$.
\end{enumerate}
\end{definition}
\noindent Figure~\ref{fig:reversal} shows splay trees for $\pi$ and $\pi'$ from Figure~\ref{fig:bigexample}.

%   -   -   -   -   -   -   -   -   -   -   -   -   -   -   -   -   -   -   -
\subsection{Reversals on trees}

\begin{figure}[]
  \begin{center}
    \includegraphics[alt={The rotate operation.},width=0.5\textwidth]{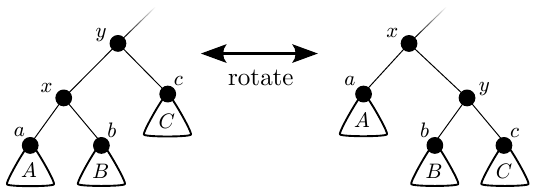}
  \end{center}
  \caption{
A rotation promotes a node above its parent, while keeping the in-order traversal intact.
}
\label{fig:rotate}
\end{figure}

The operation $splay(\pi_i)$, on a tree $T \in \T(\pi, Q)$, moves the non-root vertex $\pi_i$ to the root, producing a new tree in $\T(\pi, Q)$ having the \rev flag at the root set to \FALSE.
A splay is done through a series of \emph{rotate} operations, which are carefully chosen to make the tree self-balancing, so that $n$ splays will take $O(n \log n)$ time to perform~\cite{sleatorSelfadjustingBinarySearch1985}.
Consider a node $x$ with parent $y$, as depicted in the left panel of Figure~\ref{fig:rotate}.
The $rotate(x)$ operation promotes $x$ above its parent, while maintaining the order of the nodes visited by an in-order traversal.
%The operation is accomplished by removing two, and adding two new edges, making it easy to update the values at nodes $x$ and $y$.

The skeleton of the following lemma was implicit in our previous work, which we augment with information related to $Q = (\Qmax, \Qmin)$.
\begin{lemma} \label{lem:rotate}
A \emph{rotate} operation can be performed on a tree from $\T(\pi, Q)$ in constant time, yielding another tree from $\T(\pi, Q)$.
\end{lemma}
\begin{proof}
Without loss of generality, consider the $rotate(x)$ operation on the subtree in the left panel of Figure~\ref{fig:rotate}, creating the subtree in the right panel.

Before performing $rotate(x)$, we ensure that nodes $x$ and $y$ do not have \rev flags set to \TRUE.
If $\rev(y) = \TRUE$, we swap the order of children $x$ and $c$ while flipping the $\rev(x)$ and $\rev(c)$ values, and then we swap $\maxneg(y)$ with $\minpos(y)$ and swap $\minneg(y)$ with $\maxpos(y)$.
The product of this transformation is a tree in $\T(\pi, Q)$ since, while there is one fewer $\rev$ flag from $y$ to the root, the order of the children have been swapped, along with the extremal values.
Further, the parity of $x$ (and $c$) is even after the modification, if and only if the parity of $x$ (and $c$) was even before the modification.
If $\rev(x) = \TRUE$, then do the analogous modifications to node $x$ and its children, yielding another tree in $\T(\pi, Q)$.

Now we perform the rotate operation on a tree with $\rev(x)$ and $\rev(y)$ set to \FALSE by removing two edges and adding two edges, as depicted in Figure~\ref{fig:rotate}.
Note that the only extremal values that will change during this rotation are the values at nodes $x$ and $y$.
We set
\begin{equation*}
\maxneg(y) =
  \begin{cases}
    \max(\maxneg(b), \maxneg(c), y) & \text{if } y < 0 \land \inQmax(|y|),\\
    \max(\maxneg(b), \maxneg(c))    & \text{otherwise}.
  \end{cases}
\end{equation*}
%and
%\begin{equation*}
%\maxneg(x) =
%  \begin{cases}
%    \max(\maxneg(a), \maxneg(y), x) & \text{if } x < 0 \land \inQmax(x),\\
%    \max(\maxneg(a), \maxneg(y)) & \text{otherwise}.
%  \end{cases}
%\end{equation*}
The other extremal values for nodes $y$ and $x$ are updated in similar ways. 
\end{proof}

\begin{figure}[]
  \begin{center}
    \includegraphics[alt={Different splay trees during a reversal.},width=1.0\textwidth]{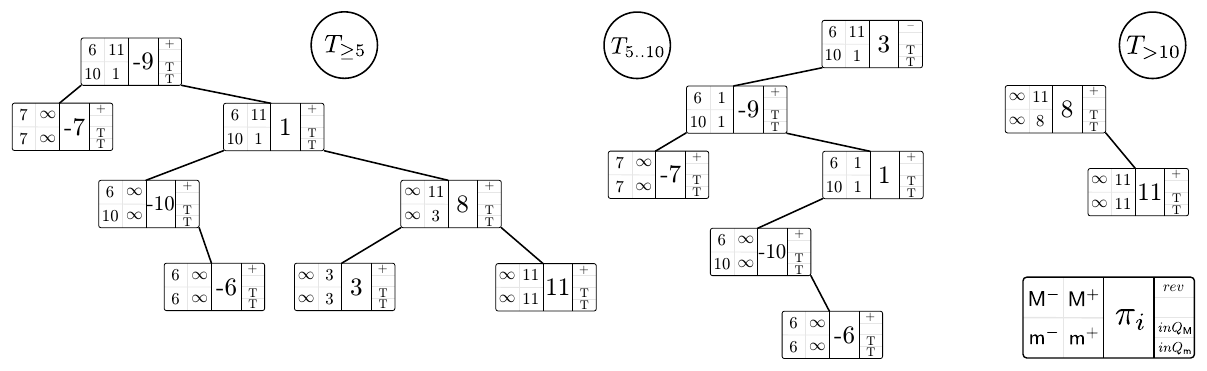}
  \end{center}
  \caption{
Subtrees created during the process of the reversal $\mu(\_7, 8) = \rho(5,10)$ on the tree from Figure~\ref{fig:splaytree}, eventually resulting in the tree of Figure~\ref{fig:reversedtree}.
$\Tge{5}$ is the right subtree of the root, after $splay(\_4)$ is done on the tree of Figure~\ref{fig:splaytree}.
$T_{5..10}$ is then obtained from \Tge{5} by doing $splay(3)$ and then splitting the right child of the root \Tgt{10}.
The tree of Figure~\ref{fig:reversedtree} is obtained by doing $splay(7)$ on $T_{5..10}$ before joining the trees back together.
}
\label{fig:revsteps}
\end{figure}

\citet{kaplanEfficientDataStructures2003} noticed that the datastructure used by \citet{chrobakDataStructureUseful1990} to reverse suffixes of travelling salesman paths could be easily adapted to reversals on permutations, by using a sequence of splay operations.
Lemma~\ref{lem:rotate} assures us that these splay operations produce a tree from the same set.
Splitting and joining splay trees at their roots are efficient for similar reasons as the rotate operation.
\begin{remark}[\citet{kaplanEfficientDataStructures2003}]
\label{rem:reversalbysplay}
A series of splay operations can update a tree $T \in \T(\pi, Q)$ through a reversal $\rho(i, j)$ to obtain a tree in $\T(\pi\rho(i,j), Q)$, in (amortized) logarithmic time.
\end{remark}
The process works in five steps.
First we
\begin{enumerate}
\item $splay(\pi_{i-1})$ and split the right child from the root, yielding trees \Tlt{i} and \Tge{i} (\ie \Tlt{i} contains the elements in $\pi_0\,\,\pi_1\, \cdots\, \pi_{i-1}$),

\item $splay(\pi_j)$ in \Tge{i} and split the right child from the root, yielding trees \Trev and \Tgt{j} (\ie \Tgt{j} contains the elements in $\pi_{j+1}\, \pi_{j+2}\, \cdots\, \pi_{n+1}$),

\item set the \rev flag at the root of \Trev,

\item $splay(last(\Trev))$ so that there is no right child of the root, and join \Tgt{j} as the right child to create $\Tge{i}'$,

\item join $\Tge{i}'$ as the right child of the root of $\Tlt{i}$,
\end{enumerate}
where $last(\Trev)$ is the node corresponding to the last element in the permutation represented by \Trev.
For an example see Figure~\ref{fig:revsteps}, which describes the intermediate states between the trees from Figure~\ref{fig:reversal} during the reversal process.

Note that, in order to make a reversal efficient, we maintain a vector of pointers from elements to their corresponding nodes.
Thus, for a good pair $(\pi_i, \pi_j)$, the node $\pi_i$ can be accessed using the vector, even when the index $i$ is not known.
To access $\pi_{i-1}$, the position $i$ can be found using the node $\pi_i$ in the search tree, before re-descending from the root to find the element at position $i-1$.
When performing a reversal for a good pair, we remove the appropriate values from  \Qmax and \Qmin, thereby ignoring the corresponding breakpoints when choosing future good pairs.

\begin{theorem} \label{thm:reversal}
%Consider a good reversal $\mu(\pi_i, \pi_j)$ on permutation $\pi$, and subsets $Q = (\Qmin, \Qmax)$ of the elements of $\pi$.
Reversal $\mu(\pi_i, \pi_j)$ can be performed on a tree from $\T(\pi, Q)$, producing a tree from $\T(\pi\mu(\pi_i, \pi_j), Q')$ in (amortized) logarithmic time, where
\begin{align*}
Q' = \big(&\Qmax \setminus \{\max(|\pi_i|, |\pi_j|)\}, \\
          &\Qmin \setminus \{\min(|\pi_i|, |\pi_j|)\}\big)
\end{align*}
%\begin{equation*}
%Q' =
%  \begin{cases}
%    \big(\Qmin \setminus \{|\pi_j|\}, \Qmax \setminus \{|\pi_i|\}\big)
%      & \text{if } |\pi_i| < |\pi_j|,\\
%    \big(\Qmin \setminus \{|\pi_i|\}, \Qmax \setminus \{|\pi_j|\}\big)
%      & \text{otherwise}.
%  \end{cases}
%\end{equation*}
\end{theorem}
\begin{proof}
First, note that the extremal values are trivial to maintain when a tree is split at the root, and when trees are joined by making one root the child of the other.

Say $\pi_i$ is negative.
After the first splay of Remark~\ref{rem:reversalbysplay}, $\pi_i$ is the root,
and we mark $\inQmax(\pi_i) = \FALSE$ when $|\pi_i| > |\pi_j|$ and $\inQmin(\pi_i) = \FALSE$ otherwise, before updating the extremal values at the root in the obvious way.
After the second splay we mark $\inQmax(\pi_j) = \FALSE$ when $|\pi_i| < |\pi_j|$ and $\inQmin(\pi_j) = \FALSE$ otherwise, and then we update the extremal values at the root in the obvious way.
Lemma~\ref{lem:rotate} ensures that \Trev is now a tree from $\T((\pi_i~\pi_{i+1}~\cdots~\pi_j), Q')$, and that
the subsequent splaying and joining of trees yields a tree in $\T(\pi\mu(\pi_i, \pi_j), Q')$.

A similar argument applies to the symmetric case of a negative $\pi_j$.
\end{proof}

% - - - - - - - - - - - - - - - - - - - - - - - - - - - - - - - - - - - - - - -
\section{Recovery}
\label{sec:recovery}

Recall from Section~\ref{sec:tannier} that when we get stuck at a permutation $\pi \cdot S$ having only positive elements, we backtrack until we find the most recent unsafe reversal $\rho_k$.
This partitions the sequence $S$ into $S_1 = \rho_1\cdots\rho_{k-1}$ and $S_2 = \rho_k\cdots\rho_{|S|}$.
Also recall from Section~\ref{sec:datastructure} that, after performing a \maxneg reversal, which creates an adjacency \adj{q}{(q+1)} (or equivalently \adj{\_(q+1)}{\_q}), we are guaranteed that $\_(q+1)$ will never be the \maxneg element and $\_q$ will never be the \minneg element in later sorting steps, so we remove $q+1$ from \Qmax and $q$ from \Qmin.
In this section, we motivate an efficient strategy for finding $\rho_k$, based on undoing reversals from $S$ until certain conditions appear on the eligible negative elements of a permutation.

When we get stuck at a permutation with all positive elements, there exist bad non-trivial components $B_1, B_2, \ldots, B_\ell$ with internal elements $I = \cup_{i=1}^\ell I_i$ framed by smallest elements $A = \cup_{i=1}^\ell a_i$ %$\{a_1, a_2, \ldots, a_\ell \}$
and greatest elements $B = \cup_{i=1}^\ell b_i$. %\{b_1, b_2, \ldots, b_\ell \}$.
At this point, sets \Qmax and \Qmin are related to these components in the following ways:
\begin{enumerate}
\item $\Qmax = I \cup B$, and
\item $\Qmin = I \cup A$.
%\item $\Qmax \cup \Qmin = I \cup A \cup B$.
\end{enumerate}
For example, say we continue sorting the $\pi$ of Figure~\ref{fig:bigexample} by doing the reversals on $\pi'$ that sort the component $\{0, \_5, \_2, \_4, \_3, \_1, 6\}$.
The remaining bad component is $\{6, 10, 9, 7, 8, 11\}$, with frame elements $6$ and $11$, implying $\Qmax = \{10, 9, 7, 8, 11\}$ and $\Qmin = \{6, 10, 9, 7, 8\}$.
%The reversal $\rho(5, 10)$ creates two new adjacencies, so $4$ and $8$ are removed from \Qmax while $3$ and $7$ are removed from \Qmin.

Denote $p_0 = \pi$, and the permutations visited in sequence $S$ as $p_j = p_{j-1}\rho_j$, for $1 \leq j \leq |S|$,
so that $p_k$ is the first permutation after the most recent unsafe reversal $\rho_k$ (\ie $k$ is maximum such that $\rho_k$ is unsafe).

The following results concern the states of \Qmax and \Qmin that exist after applying $S$ to $\pi$.
We call these sets $\Qmax^S$ and $\Qmin^S$.
Lemma~\ref{lem:allthesamesign} establishes an easy way to check that all of the bad components remain intact, when backtracking through the permutations $p_{|S|}, p_{|S|-1}, \dots, p_k$.
Lemma~\ref{lem:atleastoneneg} states that one element from the bad components of $\pi \cdot S$ must be negative in $p_{k-1} = \pi \cdot S_1$,
while Lemma~\ref{lem:atleastonegood} ensures that we can find a good pair by looking at the \maxneg and \minneg elements of the permutation $\pi \cdot S_1[\Qmax^S \cup \Qmin^S]$.

\begin{lemma}
\label{lem:allthesamesign}
%A permutation $p_j$ for $k \leq j \leq |S|$, either has no elements from $\Qmax^S$ that are negative, or $q-1$ is also negative in $p_j$ when $\_q$ is the \maxneg element in $p_j[\Qmax^S]$.
%A permutation $p_j$ for $k \leq j \leq |S|$, either has no elements from $\Qmax^S$ that are negative, or when $\_q$ is the \maxneg element in $p_j[\Qmax^S]$, $q-1$ is also negative in $p_j$ .
Consider a permutation $p_j$ for $k \leq j \leq |S|$.
%It either has no elements from $\Qmax^S$ that are negative, or when $\_q$ is the \maxneg element in $p_j[\Qmax^S]$ $q-1$ is also negative in $p_j$ .
It either has no elements from $\Qmax^S$ that are negative, or $q-1$ is negative in $p_j$ when $\_q$ is the \maxneg element in $p_j[\Qmax^S]$.
Symmetrically, $p_j$ either has no elements from $\Qmin^S$ that are negative, or $q+1$ is also negative in $p_j$ when $\_q$ is the \minneg element in $p_j[\Qmin^S]$.
\end{lemma}
\begin{proof}
By definition, $\Qmax^S$ contains all elements from each bad component in $\pi \cdot S$ except for the one in each component with smallest absolute value.
So if there is a \maxneg element $\_q$ in $p_j[\Qmax^S]$, then the smallest element of the component is $q-1$.
Since $\rho_k$ is the most recent unsafe reversal, any bad component in $p_{|S|}$ is bad in $p_j$, and all elements of such a component have the same sign, including element $q-1$.
By symmetry, the same is true for the \minneg element and $\Qmin^S$.
\end{proof}

\begin{lemma}
\label{lem:atleastoneneg}
There is at least one element from $\Qmax^S \cup \Qmin^S$ that is negative in the permutation $\pi \cdot S_1 = p_{k-1} = p_k\rho_k$.
\end{lemma}
\begin{proof}
Since $\rho_k$ is unsafe, permutation $p_k$ has a bad component that is modified by applying $\rho_k$ to it, yielding a component in $p_{k-1}$ with at least one negative element from the bad component.
The elements of the bad components of $\pi \cdot S$ are exactly $\Qmax^S \cup \Qmin^S$, so this set contains at least one negative element.
\end{proof}

\begin{lemma}
\label{lem:atleastonegood}
At least one of the \maxneg and \minneg elements from the permutation $\pi \cdot S_1[\Qmax^S \cup \Qmin^S]$ is in a good pair for $\pi \cdot S_1$.
\end{lemma}
\begin{proof}
Lemma~\ref{lem:atleastoneneg} states that at least one of the \maxneg and \minneg elements must exist in $\pi \cdot S_1[\Qmax^S \cup \Qmin^S]$.
Say that the \maxneg element $\_q$ exists.
Then by definition $\_q$ is an element from a bad component $B_i$ of $\pi \cdot S$ other than the frame element $a_i$, and the element $q-1 \geq a_i$ must be positive.
Thus, $(q-1, \_q)$ is a good pair. %, due to Lemma~\ref{lem:anygood}.
By symmetry the same holds for the \minneg element, when it exists.
\end{proof}

These lemmas point to a strategy for efficient detection of the most recent unsafe reversal, once stuck at permutation $\pi \cdot S$ with allowed pair $Q^S = (\Qmax^S, \Qmin^S)$.
Start with a tree in $\T(\pi \cdot S, Q^S)$, and undo reversals $\rho_{|S|}\rho_{|S|-1}\cdots\rho_k$ until the following check (or the symmetric check on the \minneg element) passes: $\maxneg(r) = q \neq -\infty$ for the root $r$ of a tree from the set $\T(\pi \cdot \rho_{|S|}\rho_{|S|-1}\cdots\rho_k, Q^S)$, and $q-1$ is positive.

% - - - - - - - - - - - - - - - - - - - - - - - - - - - - - - - - - - - - - - -
\section{The algorithm}
\label{sec:algorithm}

We present Algorithm~\ref{alg:sort_permutation}, which is a version of Algorithm~\ref{alg:sort_graph} that uses the tree from Section~\ref{sec:datastructure} to find and detect good reversals.
The structure of the algorithms are almost identical, the only difference being that identity pairs that have yet to form an adjacency are maintained in the \inQmax and \inQmin values on the tree, rather than in the set $Q$.

\alg{AllBad} tests if there exists a good pair in $T$.
It is implemented by checking the \maxneg and \minneg values at the root according to the reasoning of Lemmas~\ref{lem:allthesamesign} and \ref{lem:atleastoneneg}:
if the \maxneg value at the root is $\_q$ and the sign of element $q-1$ is positive, or if the \minneg value is $\_q$ and the sign of element $q+1$ is positive, then we return \FALSE.
Otherwise we return \TRUE.

Lemma~\ref{lem:atleastonegood} ensures that \algarg{FindGood}{T} can use the \maxneg and \minneg value at the root of $T$ in each successive call to \alg{DoGood}.

%The recursive call to \alg{SortGoodTree} is performed at most $n$ times, since by Lemma~\ref{lem:anygood} $S$ will contain at least two reversals, and since each of the calls to \alg{SortGoodTree} is independent from the previous.
The recursive call to \alg{SortGoodTree} is performed at most $n$ times for the same reasons that \alg{SortGraph} is called at most $n$ times in Section~\ref{sec:sortingOG}.
By Theorem~\ref{thm:reversal} each reversal takes $O(\log n)$ time to perform, and each identity pair is associated to at most two reversals; one the applies the reversal, and one that may undo it.
The running time of the algorithm, therefore, is $O(n \log n)$.

\begin{algorithm}[] %H
\caption{Sort a signed permutation $\pi$.}
\label{alg:sort_permutation}
\begin{algorithmic}[0]   %1 for line numbers

\Statex $\triangleright$ Compute sequence of reversals that sorts permutation $\pi$.
\Procedure{SortSignedPermutation}{$\pi$}
  \Let{$\pi, R$}{\algarg{TransformBad}{\pi}}
    \Comment{make bad components good \cite{baderLinearTimeAlgorithmComputing2001,bergeronCommonIntervalsSorting2002}}
  \Let{Q}{\algarg{GetGoodPairs}{\pi}} \Comment{$Q = (\Qmax, \Qmin)$}
  \Let{T}{\algarg{MakeTree}{\pi, Q}}
  \Let{$T, S$}{\algarg{SortGoodTree}{T}}
  \Return{$R \cdot S$}
\EndProcedure

\Statex
\Statex $\triangleright$ Find an order of good pairs that sorts $T$.
\Procedure{SortGoodTree}{$T$}
  \Let{$T, \Sfront$}{\algarg{DoGood}{T}}
  \State \Sback is an empty sequence
  \Comment{$f$ for ``front'' and $b$ for ``back''}
  \While{\Sfront is not empty} \Comment{good pairs remain}
    \Let{$T, S_1, S_2$}{\algarg{Recover}{T, \Sfront}}
      \Comment{$S_1 \cdot S_2 = \Sfront$}
    \Let{\Sfront}{$S_1$}
    \If{\Sfront is empty} \Comment{\alg{Recover} found no good pairs}
       \Return{$T, S_2 \cdot \Sback$}
    \EndIf
    \Let{$T, S'$}{\algarg{SortGoodTree}{T}}
    \If{$even(|S'|)$}
       \Let{\Sback}{$S' \cdot S_2 \cdot \Sback$}
    \Else
       \Let{\Sback}{$S' \cdot S_2^- \cdot \Sback$}
       \Comment{$S_2^-$ is $S_2$ without the first vertex}
    \EndIf
  \EndWhile
  \Return{$T, \Sfront \cdot \Sback$}
\EndProcedure

\Statex
\Statex $\triangleright$ Do \maxneg or \minneg reversals for elements still in \Qmax or \Qmin.
\Procedure{DoGood}{$T$}
  %\Let{$S$}{()}
  \State $S$ is an empty sequence
  \While{\algarg{HasNegative}{T}} \Comment{\maxneg or \minneg exists at root}
    \Let{$\rho$}{\algarg{FindGood}{T}}
      \Comment{via the \maxneg or \minneg value at root}
    \Let{$T$}{$T \rho$} \Comment{via Theorem~\ref{thm:reversal}, updating $Q$ values}
    \State $append(S, \rho)$
  \EndWhile
  \Return{$T, S$}
\EndProcedure

\Statex
\Statex $\triangleright$ Get longest prefix of $S$ such that $T$ has good pair (by undoing reversals).
\Procedure{Recover}{$T, S$}
  %\Let{$S_2$}{()}
  \State $S_2$ is an empty sequence
  \While{\algarg{AllBad}{T}}
    \Let{$\rho$}{$pop(S)$} \Comment{$pop()$ removes from the end}
    %\State $prepend(\rho, S_2)$
    \Let{$S_2$}{$\rho \cdot S_2$}
    \Let{$T$}{$T \rho$} \Comment{via Remark~\ref{rem:reversalbysplay}, without updating $Q$ values}
  \EndWhile
  \Return{$T, S, S_2$}
\EndProcedure

\end{algorithmic}
\end{algorithm}

% - - - - - - - - - - - - - - - - - - - - - - - - - - - - - - - - - - - - - - -
\section{Conclusion}
\label{sec:concusion}

Our algorithm is the product of adapting our previous datastructure, suitable for finding the maximum and minimum negative elements of a permutation, to the framework of \citet{tannierAdvancesSortingReversals2007}.
The first adaptation was the addition of the notion of ``eligible'' elements to our datastructure.
The second was the proof that maximum and minimum negative elements can be used to detect bad reversals.
In order to effectively use these adaptations, it required a re-imagining of the \citet{tannierAdvancesSortingReversals2007} results.

While the discovery of a faster algorithm for \SSBR might be surprising, the gap between our upper bound and the trivial lower bound remains open.
Eliminating the gap would likely require a fresh perspective, and a dose of ingenuity.

% - - - - - - - - - - - - - - - - - - - - - - - - - - - - - - - - - - - - - - -
\section*{Acknowledgements}

We would like to thank the sorting by reversals reading group, composed of Severine Berard, Annie Chateau, Celine Mandier, Jordan Moutet, and Pengfei Wang, for their participation and engaging discussions.
We would also like to thank Anne Bergeron, Askar Gafurov, and Bernard Moret for their helpful remarks during the preparation of this manuscript.

\bigskip
\noindent Dedicated to the memory of my friend Yu Lin.

\newpage
\bibliography{bibl}{}
\bibliographystyle{plainnat}

\end{document}